\documentclass[a4paper]{jpconf}
\usepackage{graphicx}
\usepackage{amssymb}
\begin{document}
\title{Very Hight Energy Observationa of Shell-Type Supernova Remnants with \\ SHALON Mirror Cherenkov
Telescopes}

\author{V.G. Sinitsyna, V.Y. Sinitsyna}

\address{P.N. Lebedev Physical Institute, Leninsky pr. 53, Moscow, Russia}

\ead{sinits@sci.lebedev.ru}

\begin{abstract}
The investigation of VHE gamma-ray sources by any methods, including
mirror Cherenkov telescopes, touches on the problem of the cosmic
ray origin and, accordingly, the role of the Galaxy in their
generation. The SHALON observations have yielded the results on
Galactic shell-type supernova remnants (SNR) on different evolution
stages. Among them are: SNRs Tycho's SNR, Cas A, IC 443,
$\gamma$Cygni SNR and classical nova GK Per (Nova 1901). For each of
SNRs the observation results are presented with spectral energy
distribution by SHALON in comparison with other experiment data and
images by SHALON. The collected experimental data have confirmed the
prediction of the theory about the hadronic generation mechanism of
very high energy 800 GeV - 100 TeV $\gamma$-rays in Tycho's SNR, Cas
A and IC443.
\end{abstract}

\section*{Introduction}
The detection of TeV gamma-rays from the number of galactic sources
was the major achievement of gamma-astronomy and is providing new
insights into the cosmic ray emission mechanisms and their sources.
The hypothesis that Supernova Remnants (SNRs) are unique candidates
for cosmic-ray sources has been prevalent from the very beginning of
cosmic-ray physics. The electron component of Cosmic Rays is well
visible on the SNR emission through the wide range of
electromagnetic spectrum from radio to low energy gamma-ray
emission, but the information on the nuclear Cosmic Ray component in
SNRs can only be obtained from high and very high energy gamma-ray
observations. So, gamma-astronomic observations can help to solve
the cosmic ray origin. We present the observation results by SHALON
telescope of shell type Galactic supernova remnants: Tycho's SNRs,
Cas A, $\gamma$Cygni SNR and IC 443 as well as the classical nova GK
Per that is on the earliest SNR evolution stage.

\section*{GK Per (Nova 1901)}
Nova Persei 1901 (GK Per) is one of the most extensively observed
and studied classical nova shells over the entire electromagnetic
spectrum. The optical data \cite{GKPer_Opt} are demonstrated
interaction between the nova ejecta and the ambient gas.
Furthermore, remnant of nova is detected at radio energies with the
Very Large Array (VLA) as a source of nonthermal, polarized radio
emission \cite{GKPer_VLA}. The results of these observations show
the existence of shocked interstellar material. The X-ray shell
around GK Per was first discovered with the ROSAT experiment and
then it has been observed by Chandra telescope \cite{GKPer_Chandra}.
In particular, with Chandra observations, the X-ray emission of the
same electron population has been detected as the extension from the
radio wavelengths. The detection of the X-rays from the supernova
remnant shell which are primarily due to bremsstrahlung of shock
accelerated relativistic electrons, supposed the detection of
gamma-ray emission originated from $\pi^\circ$- decay, secondary
pp-interactions \cite{GKPer_Berezhko} as well as possible
contribution emission produced via Inverse Compton scattering
\cite{GKPer_Berezhko}. Chandra X-ray data shows that, the nova
remnant of GK Per could be a younger remnant that will resemble
older SNRs like IC 443 $((3 \div 30)\times10^3$ year) which interact
with molecular clouds.

\begin{figure}[!]
\centering
\includegraphics[width=2.55in]{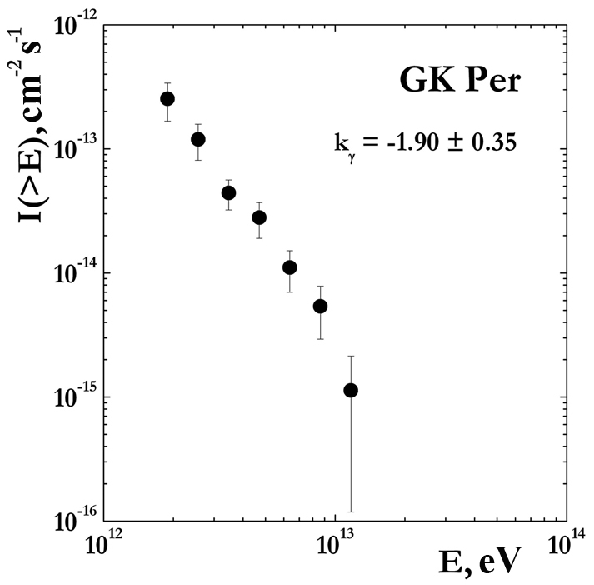}
\includegraphics[width=2.55in]{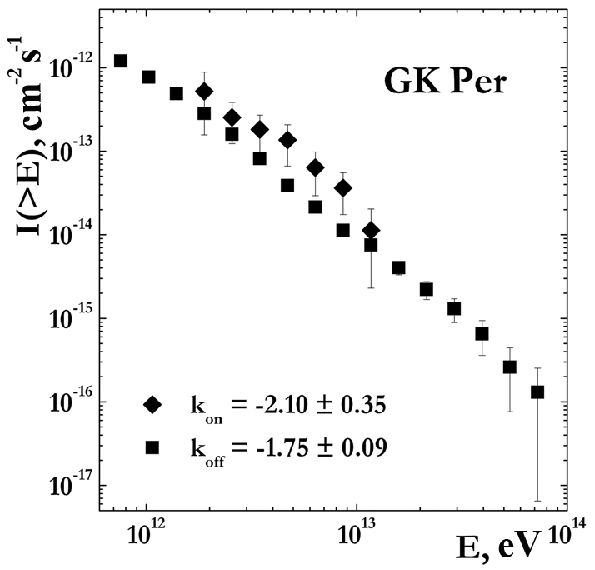}\\
\caption{GK Per (Nova 1901) characteristics: {\bf{left}} - Integral
gamma-ray spectrum with a power-law index $k_{\gamma} = -1.90\pm
0.35$; {\bf{right}} - The event spectrum from the source with
background with index of $k_{ON} = -2.10\pm 0.35$ and spectrum of
background events observed simultaneously with the object with index
$k_{OFF} = -1.75 \pm 0.09$ } \label{on_off}
\end{figure}

\begin{figure}[!]
\centering
\includegraphics[width=3.15in]{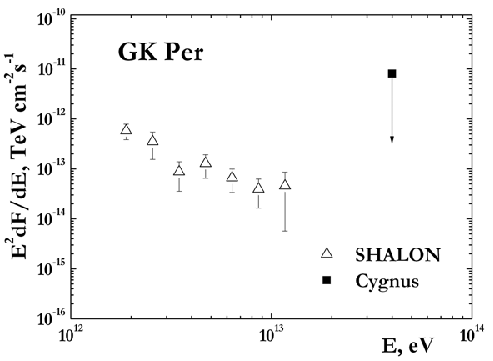}
\includegraphics[width=2.85in]{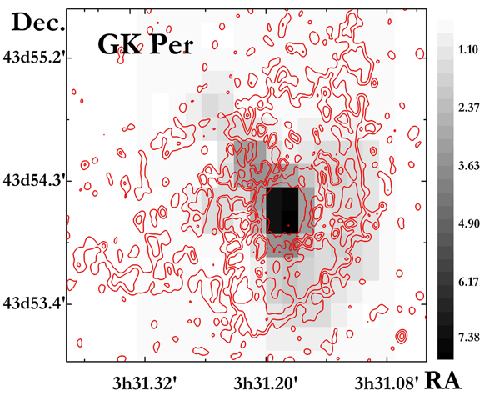}\\\vspace{-0.1cm}
\caption{GK Per (Nova 1901) characteristics: {\bf{left}} - Spectral
energy distribution from GK Per (Nova 1901) obtained by SHALON;
 {\bf{right}} - The image of GK Per in the energy range 800 GeV - 40 TeV
by SHALON the contours indicate Chandra X-ray image of GK Per [3].}
\label{SNR_GK}
\end{figure}

In accordance with program on long-term studies of metagalactic
gamma-ray sources, fifteen-year-long observations of the central
galaxy in the Perseus cluster, NGC 1275, are being carried out in
the SHALON experiment
\cite{Sinitsyna2014,Sinitsyna1996,Sinitsyna2011,Sinitsyna2001,Sinitsyna2013,SNR_Como}.
During the observations of NGC 1275 the SHALON field of view
contains the source of nonthermal radio and X-ray emission GK Per
(Nova 1901) of classical nova type as it located at $~3^\circ$ North
from NGC 1275. So due to the large telescopic field of view $>
8^\circ$ the observations of NGC 1275 is naturally followed by the
tracing of GK Per. GK Per as a source accompanying to NGC 1275 was
observed with SHALON telescope at the period from 1996y to 2012y for
a total of 111 hours during the clear moonless nights at zenith
angles from $5^\circ$ to $35^\circ$. The observations were performed
using the standard (for SHALON) technique of obtaining information
about the cosmic-ray background and gamma-ray-initiated showers in
the same observing session. The SHALON method of selecting
$\gamma$-ray showers from background cosmic-ray showers allow to
reject 99.92\% of the background showers
\cite{Sinitsyna2014,Sinitsyna1996,Sinitsyna2011}. The $\gamma$-ray
source associated with the GK Per was detected above 2 TeV by SHALON
with a statistical significance $9.2\sigma$ determined by Li\&Ma
method \cite{LiMa}. The signal significance for this SNR is less
then one for the source with similar flux and spectrum index
obtained in the same observation hours because of less collection
field of view relative to the standard procedure of SHALON
experiment \cite{Sinitsyna2014,Sinitsyna1996,Sinitsyna2011}. The
corrections for the effective field of view were made to calculate
source flux and energy spectrum. The average integral flux at
energies above 2 TeV for GK Per is $I_{GK Per} = (2.9 \pm 1.3)\times
10^{-13} cm^{-2}s^{-1}$ (Fig. \ref{on_off}, left). Figure 1b
presents the spectra of the ON- and OFF-events needed to extract the
gamma-ray spectrum from GK Per. The gamma-ray spectrum of GK Per
(Fig. \ref{on_off}, left) is obtained by subtracting the spectrum of
the background events recorded simultaneously with the sources
observations, $I_{OFF} \propto E^{k_{OFF}}$, from the spectrum of
the events arrived from the source position, $I_{ON} \propto
E^{k_{ON}}$ (Fig. \ref{on_off}, right). Taking into account the
proximity to a nearby source NGC 1275, we made the observation data
procession first associated with NGC 1275 and then with GK Per. We
found that 1\% of showers are common for the both sources. After the
detailed analysis of arrival direction of these showers and angular
distance \cite{Sinitsyna2014}, less than 0.5\% of NGC 1275 showers
were recognized to be GK Per showers. This didn't change the average
flux of NGC 1275 form \cite{Sinitsyna2014,Sinitsyna2013,SNR_Como}.
The energy spectrum of $\gamma$-rays in the observed energy region
from 2 to 15 TeV is well described by the power law $F(E > 2$
TeV$)\propto E^k_{\gamma}$ , with $k_{\gamma} = -1.90\pm0.35$ (see
Fig. \ref{on_off}, left). The spectral energy distribution and the
image of GK Per at TeV-energies by SHALON are shown in Figure
\ref{SNR_GK}. The analysis of $\gamma$-ray shower arrival direction
revealed the main TeV-emission region coinciding with the position
of central source of GK Per and the weak emission of shell, that is
also observed in X-ray by Chandra \cite{GKPer_Chandra}
(Fig.\ref{SNR_GK}, right).
\begin{figure}[t!]\vspace{-1.1cm}
\centering \vspace{-1.1cm}
\includegraphics[width=2.95in]{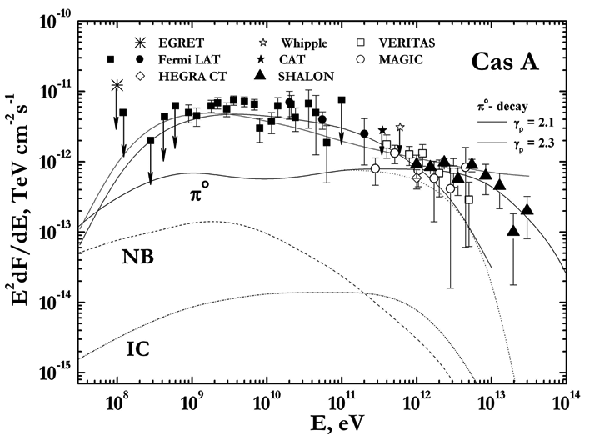}
\includegraphics[width=2.64in]{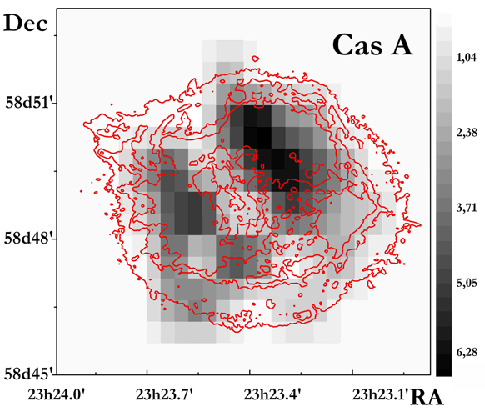}\\\vspace{-0.15cm}
\includegraphics[width=2.95in]{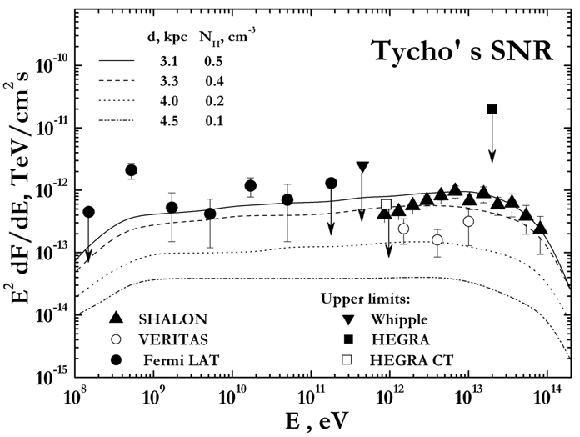}
\includegraphics[width=2.64in]{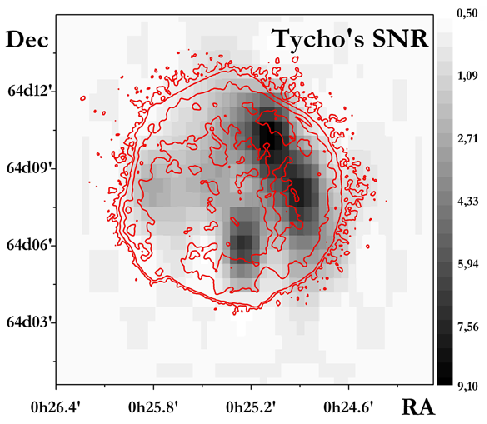}\\\vspace{-0.05cm}
\caption{Characteristics of shell-type supernova remnants Cas A and
Tycho's SNR. {\bf{left}} - Spectral energy distributions of high-
and very high energy $\gamma$-ray emission by SHALON
($\blacktriangle$) in comparison with other experiments (see text).
NB - nonthermal Bremsstrahlung $\gamma$-ray energy flux; IC -
Inverse Compton $\gamma$-ray energy flux \cite{CasA_Berezhko}; d -
source distance; $N_H$ - interstellar medium density. {\bf{right}} -
images of SNRs at energies $> 0.8$ TeV by SHALON (grey scale); red
contours are the X-ray emission by Chandra. }\label{SNR1}
\end{figure}

\section*{Cas A supernova remnant (1680 year)}
Cas A is a youngest of historical supernova remnant in our Galaxy.
Its overall brightness across the electromagnetic spectrum makes it
a unique object for studying high- and very high- energy phenomena
in SNRs. Cas A was observed with SHALON telescope during the 68
hours in period of 2010 - 2013 yy \cite{Sinitsyna2011}. The
gamma-ray source associated with the SNR Cas A was detected above
800 GeV with a statistical significance \cite{LiMa} of 16.1$\sigma$
with a gamma-quantum flux above 0.8 TeV of $(0.64 \pm
0.10)\times10^{-12} cm^{-2} s^{-1}$. The energy spectrum of
gamma-rays in the observed energy region from 0.8 TeV to 30 TeV is
well described by the power law with exponential cutoff,
$I(>E_{\gamma}) = (0.64 \pm 0.10)\times10^{-12}\times E_{\gamma}^
{-0.91\pm0.11}\times exp(-E_{\gamma}/10,3 TeV)$ (Fig. \ref{SNR1},
left). Figure \ref{SNR1} right presents Chandra X-ray image of Cas A
(lines) \cite{CasA_Chandra} in comparison TeV structure in energy
range of 0.8 - 30 TeV by SHALON.

Two favored scenarios in which the gamma-rays of 500 MeV - 10 TeV
energies are emitted in the shell of the SNR like Cas A are
considered. The gamma-ray emission could be produced via Inverse
Compton scattering and by accelerated cosmic ray hadrons through
interaction with the interstellar gas and then $\pi^{\circ}$- decay.
Figure \ref{SNR1}, left presents spectral energy distribution of the
gamma-ray emission from Cas A by SHALON ($\blacktriangle$) in
comparison with theoretical predictions
\cite{CasA_Fermi,CasA_Berezhko} and other experimental data: Fermi
LAT \cite{CasA_Fermi}, HEGRA \cite{CasA_Hegra}, MAGIC
\cite{CasA_Magic}, VERITAS \cite{CasA_Veritas}; upper limits of
EGRET \cite{CasA_Egret}, CAT \cite{CasA_Cat}, Whipple
\cite{CasA_Whipple}. Solid lines show the very high energy gamma-ray
spectra of hadronic origin \cite{CasA_Fermi,CasA_Berezhko}. The
calculation results of leptonic model in assumption of two magnetic
field values are also considered in \cite{CasA_Fermi}. It was shown
that leptonic model with B = 0.3 mG predicts a 5 - 8 times lower
gamma-ray flux than the observed; the model with B = 0.12 mG, which
can broadly explain the observed GeV flux predicts the TeV spectrum
with cut-off energy about 10 TeV.

The detection of gamma-ray emission at 5 - 30 TeV and the hard
spectrum below 1 TeV would favor the $\pi^\circ$-decay origin of the
gamma-rays in Cas A SNR.

\section*{Tycho's Supernova Remnant  (1572 year)}
Tycho's SNR originated from the Ia type supernova which exploded in
1572 year. The high quality X-ray image of Tycho's SNR by Chandra
shows an expanding bubble of debris inside a more rapidly moving
shell of extremely high energy electrons. The supersonic expansion
of the stellar debris has created two X-ray emitting shock waves -
one moving outward into the interstellar gas, and another moving
back into the debris. Such the character of displacement of the
shock and the contact discontinuity surfaces makes the cosmic ray
acceleration at the supernova shock very efficient.

The observations of Tycho supernova remnant are carried out by
SHALON telescope in the period of 1996 - 2010 years
\cite{Sinitsyna1996,Sinitsyna2011,Sinitsyna2001,Sinitsyna2013,SNR_Como}.
Tycho's SNR has been detected by SHALON at energies above 0.8 TeV
with a statistical significance 17$\sigma$ determined by Li\&Ma
method \cite{LiMa}. The integral gamma-ray flux above 0.8 TeV was
estimated as $(5.2\pm0.4)\times 10^{-13} cm^{-2}s^{-1}$. The energy
spectrum of $\gamma$-rays in the observed energy region from 0.8 to
80 TeV by SHALON is well described by the power law with exponential
cutoff $I(>E_{\gamma}) = (0.41 \pm 0.05) \times 10^{-12} \times
E_{\gamma} ^{-0.93\pm0.09}\times exp(-E_{\gamma}/35 TeV)$. The image
of Tycho's SNR at energies 0.8 - 80 TeV by SHALON telescope is also
shown in Fig. 3. Recently, Tycho's SNR was also confirmed with
VERITAS in observations of 2008 - 2010 and 2011years
\cite{Tycho_Veritas}. The $\gamma$-ray emission from Tycho'SNR was
detected with Fermi LAT in the range 100 MeV - 300 GeV
\cite{Tycho_Fermi}. According to model \cite{Tycho_Berezhko1} the
expected flux of $\gamma$-quanta from $\pi^\circ$- decay, $F\propto
E^{-1}$ , extends up to $>30$ TeV, while the flux of $\gamma$-rays
originated from the Inverse Compton scattering has a sharp cutoff
above the few TeV. So the detection of $\gamma$-rays with energies
10 - 80 TeV from Tycho's SNR by SHALON is an evidence of their
hadronic origin.

Due to the high-quality observations from XMM-Newton and Chandra the
fundamental properties of the X-ray emission in Tycho, as a SN
explosion energy $E_{SN} = 1.2\times10^{51}$ erg, which are
necessary to model calculations are available. The distance
determinations for Tycho's SNR have varied from 2.0 - 2.8 kpc to 3.1
- 4.5 kpc \cite{Tycho_Berezhko2}. In order to find a constraint on
the source parameters as distance d and the interstellar medium
density $N_H$, we compare the model resulting $\gamma$-ray spectral
energy distribution \cite{Tycho_Berezhko2} with one from
$\gamma$-ray observations at TeV energies. The additional
information about parameters of Tycho's SNR can be obtained in frame
of nonlinear kinetic model \cite{Tycho_Berezhko1,Tycho_Berezhko2} if
the TeV -quantum spectrum of SHALON telescope is taken into account:
a source distance 3.1 - 3.3 kpc and an ambient density $N_H = 0.4 -
0.5 cm^{-3}$ and the expected $\pi^\circ$-decay $\gamma$-ray energy
spectrum extends up to about 100 TeV (Fig. \ref{SNR1}, left). The
same parameters have obtained in \cite{Tycho_Blinnikov} calculations
of structures visible by Chandra at X-ray energies (see red contours
in Fig. \ref{SNR1}, right).\vspace{-0.4cm}
\begin{figure}[t!]\vspace{-1.1cm}
\centering \vspace{-1.1cm}
\includegraphics[width=2.96in]{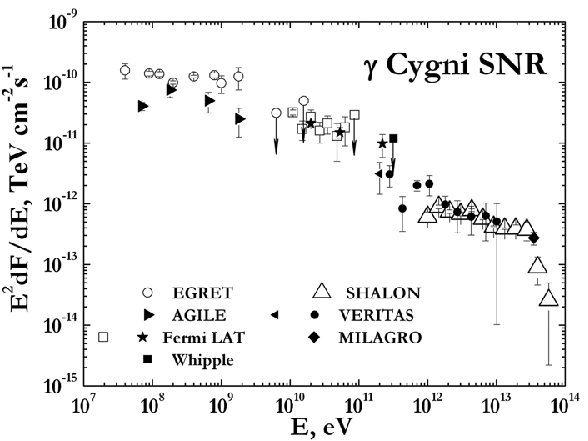}
\includegraphics[width=2.71in]{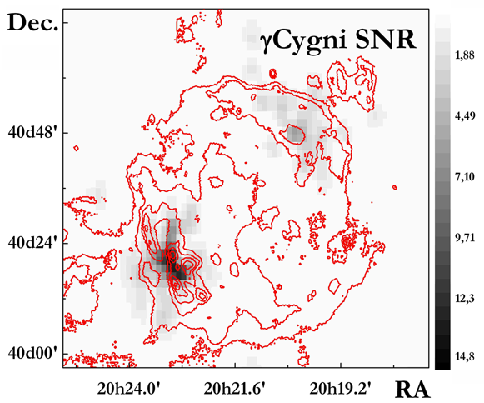}\\\vspace{-0.0cm}
\includegraphics[width=2.96in]{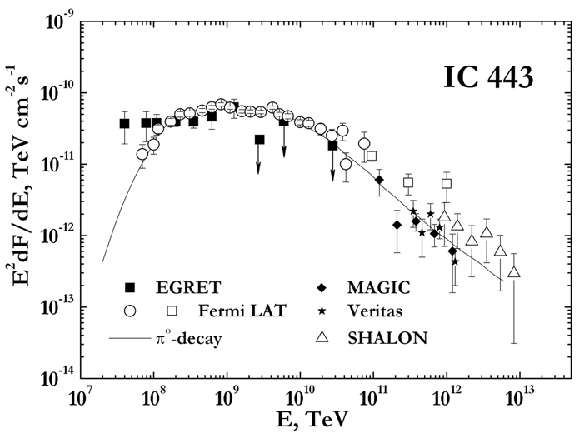}
\includegraphics[width=2.65in]{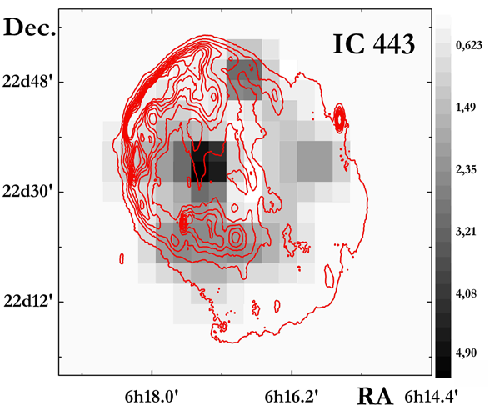}\\\vspace{-0.3cm}
\caption{Characteristics of shell-type supernova remnants
$\gamma$Cygni SNR and IC 443. {\bf{left}} - Spectral energy
distributions of high- and very high energy $\gamma$-ray emission by
SHALON ($\triangle$) in comparison with other experiments (see
text). {\bf{right}} - images of SNRs at energies $> 0.8$ TeV by
SHALON (grey scale); red contours are radio structure by Canadian
Galactic Plane Survey. }\label{SNR2}
\end{figure}
\section*{$\gamma$Cygni SNR (age $\sim(5 \div 7)\times10^3$ years)}
$\gamma$Cygni SNR is a shell-type supernova remnant at a distance of
1,5 kpc and with the observed diameter of $~1^\circ$. The shell-like
features are known in radio- and X-ray energy regions
\cite{GamCyg_XRay}. $\gamma$Cygni SNR is older then Cas A and
Tycho's SNR, its age is estimated as 5000 - 7000 years.
\cite{GamCyg_XRay,GamCyg_XRay1} and its supposed to be and in an
early phase of adiabatic expansion. The observations of different
age supernova remnants can help to reveal the mechanisms of very
high energy cosmic ray acceleration in the SNRs up to energies
$10^{15}$ eV.

$\gamma$Cygni SNR as a source accompanying to Cyg X-3 is
systematically studied with SHALON telescope since 1995y up to now.
The $\gamma$-ray source associated with the $\gamma$Cygni SNR was
detected by SHLAON \cite{Sinitsyna2013,SNR_Como} above 800 GeV with
a statistical significance \cite{LiMa} of 14$\sigma$. The average
integral gamma-flux above 0.8 TeV: $I_{\gamma Cygni SNR} = (1.27\pm
0.11)\times10^{-12} cm^{-2}s^{-1}$. The signal significance for this
SNR is less then one for the source with similar flux and spectrum
index obtained in the same observation hours because of less
collection field of view relative to the standard procedure of
SHALON experiment \cite{Sinitsyna2014,Sinitsyna1996,Sinitsyna2011}.
The corrections for the effective field of view were made to
calculate source flux and energy spectrum. The energy spectrum of
$\gamma$-rays in the observed energy region from 800 GeV to 50 TeV
is well described by the power law with exponential cutoff, $(1.12
\pm 0.11)\times10^{-12}\times E_{\gamma}^{-0.93\pm 0,09}\times
exp(-E_{\gamma}/20TeV)$ \cite{Sinitsyna2013,SNR_Como} (Fig.
\ref{SNR2}, left). The image of $\gamma$Cygni SNR at energies 0.8 -
50 TeV by SHALON is presented in Fig. \ref{SNR2}, right in
comparison with radio structure by Canadian Galactic Plane Survey
(CGPS) (see lines).  The analysis of TeV gamma-ray arrival
directions reveal two emission regions in $\gamma$Cygni SNR: the
main at the South-East part of SNR shell and second one at North.
The main contribution of energy flux gives the SE region of SNR
shell. Also, TeV $\gamma$-ray emission regions correlate with the NW
and SE parts of the shell visible in the radio energies by CGPS. The
VERJ2019+407 source was detected at 200GeV by VERITAS
\cite{GamCyg_Veritas} correlated with the position of northern part
of $\gamma$Cygni SNR shell. In Figure 3 the spectral energy
distribution of the $\gamma$-ray emission from $\gamma$Cygni SNR by
SHALON is presented in comparison with experiment data from EGRET
\cite{CasA_Egret}, AGILE \cite{GamCyg_Agile}, Fermi LAT
\cite{GamCyg_Fermi1,GamCyg_Fermi2,GamCyg_Fermi3}, VERITAS
\cite{GamCyg_Veritas,GamCyg_Veritas1} è MILAGRO
\cite{GamCyg_Milagro}.

\section*{IC 443 supernova remnant (age $\sim(3 \div 30)\times10^3$ years)}
The IC 443 is well known for its radio, optical, X-ray, and MeV-TeV
energy  $\gamma$-ray emissions. IC 443 is a shell-type SNR and it
has an angular extent of 45' in the radio energies with a complex
shape consisting of two half-shells with different radii. The age of
IC 443 remains uncertain, with various estimates placing it in the
range $(3\div 4)\times10^3$ yr., but other analysis  indicate that
it is older $(20\div 30)\times10^3$ yr. IC 443 is one of the best
candidates for the investigation of the connection among SNRs,
molecular clouds and high- and very high energy $\gamma$-ray
sources. The close placement of the dense shocked molecular clouds
and GeV-TeV $\gamma$-ray emission regions \cite{IC443_Fermi}
detected by EGRET, Fermi LAT, MAGIC and VERITAS suggests that IC 443
can be considered as a candidate to the hadronic cosmic-ray source.

IC 443 was detected by SHALON with the integral flux above 0.8TeV
$(1.69\pm0.58)\times10^{-12} cm^{-2} s^{-1}$  \cite{SNR_Como} with a
statistical significance of 9.7$\sigma$ \cite{LiMa}. The integral
energy spectrum of IC 443 can be approximated by the power law with
index $k_{\gamma} = -1.94\pm 0.16$. The favored scenario in which
the $\gamma$-rays of 100 MeV - 7 TeV energies are emitted in the
shell of the IC443 SNR is $\pi^\circ$-decay which produced in the
interactions of the cosmic rays with the interstellar gas. Inverse
Compton scattering can not explain the observed IC 443 $\gamma$-ray
emission as there is no bright source of seed photons in the region
of the IC 443. The spectral energy distribution of the $\gamma$-ray
emission from IC443 by SHALON ($\triangle$) in comparison with other
experiment data EGRET, Fermi LAT, MAGIC, VERITAS \cite{IC443_Fermi}
and with theoretical predictions is shown in Figure Fig. \ref{SNR2},
right. Solid line shows the very high energy $\gamma$-ray spectra of
hadronic origin. The image of IC443 at TeV energies by SHALON is
presented in Figure 3 in comparison with radio structure by CGPS
(see lines). The analysis of arrival directions of $\gamma$-rays
with energies 800 GeV - 7 TeV reveal the correlation of TeV
$\gamma$-ray emission maxima with MeV-GeV emission observed by Fermi
LAT \cite{IC443_Fermi}, also TeV gamma-ray emission of South and
South-West parts of IC 443 shell correlated with the position of
swept out dense molecular cloud (Fig. \ref{SNR2}, right).

\section*{Conclusion}
The observation results of Galactic shell-types supernova remnants
on different evolution stages GK Per (Nova 1901), Cas A, Tycho's
SNR, $\gamma$Cygni SNR and IC 443 by SHALON mirror Cherenkov
telescope are presented. The TeV gamma-ray emission of classical
nova GK Per, that could be a shell-type supernova remnant on early
evolution stage, was detected for the first time by SHALON. Also,
very high energy gamma-rays from the shell of GK Per, visible in the
X-rays, were detected with SHALON experiment for the first time. The
experimental data have confirmed the prediction of the theory about
the hadronic generation mechanism of very high energy  $\gamma$-rays
in Tycho's SNR, Cas A and IC443.

\end{document}